\documentclass[a4paper,11pt]{article}

\usepackage{graphicx}
\usepackage[a4paper,margin=2.7cm]{geometry}
\usepackage[pagebackref=false,colorlinks,linkcolor=blue,citecolor=red]{hyperref}
\usepackage{cite}
\usepackage{rotating}
\usepackage{xcolor}
\usepackage{sectsty}

\begin{document}

\title{Ocular Brachytherapy Dosimetry for $^{103}Pd$ and $^{125}I$ in The Presence of Gold Nanoparticles: \\
Monte Carlo Study}
\author{S.Asadi$ ^{\,1*}$, M.Vaez-zadeh$ ^{\,1}$, M.Vahidian$ ^{\,1}$, M.Marghchouei$ ^{\,1}$,  S.Farhad Masoudi$ ^{\,1}$\\
\scriptsize $ ^{\,1}$ Department of Physics, K.N. Toosi University of Technology, Tehran, Iran.\\
}

\date{}

\maketitle
%\sectionfont{\color{red}}  % sets colour of sections
\begin{abstract}\label{abs}

The aim of the present Monte Carlo study is to evaluate the variation of energy deposition in healthy tissues in the human eye which is irradiated by brachytherapy sources in comparison with the resultant dose increase in the gold nanoparticle(GNP)-loaded choroidal melanoma. The effects of these nanoparticles on normal tissues are compared between $^{103}Pd$ and $^{125}I$ as two ophthalmic brachytherapy sources. Dose distribution in the tumor and healthy tissues have been taken into account for both mentioned brachytherapy sources. Also, in a certain point of the eye, the ratio of the absorbed dose by the normal tissue in the presence of GNPs to the absorbed dose by the same point in the absence of GNPs has been calculated. In addition, differences observed in the comparison of simple water phantom and actual simulated human eye in presence of GNPs are also a matter of interest that have been considered in the present work. The results show that the calculated dose enhancement factor in the tumor for $^{125}I$ is higher than that of the $^{103}Pd$. The difference between the eye globe and the water phantom is more obvious for $^{125}I$ than that of the $^{103}Pd$ when the ophthalmic dosimetry is done in the presence of GNPs. Whenever these nanoparticles are utilized in enhancing the absorbed dose by the eye tumor, the use of $^{125}I$ brachytherapy source will greatly amplify the amount of dose enhancement factor (DEF) in the tumor site without inflicting much damage to healthy organs, when compared to the $^{103}Pd$ source. Furthermore in Monte-Carlo studies of eye brachytherapy, more precise definition of the eye phantom instead of a water phantom will become increasingly important when we use $^{125}I$ as opposed to when $^{103}Pd$ is considered.
$^{*}$Corresponding author: S.Asadi,E-mail address: s\_asadi@sina.kntu.ac.ir

\end{abstract}

\vspace*{4cm}
\renewcommand{\thesection}{\Roman{section}}
\section{Introduction}

Uveal Melanoma is one of the primary ocular cancerous tumors which arises within the eyeball in the Uvea involving the Iris, Ciliary body or Choroid. In the majority of cases, ocular melanoma develops slowly from the pigmented cells of the choroid which are named choroidal melanoma. Although this kind of cancer is rare, it is the most common eye cancer in people who are middle-aged or older ~\cite{a,b1,n,c1}.\\ Regarding the size and location of the tumor and also the rate of its progress, treatment of the choroidal melanoma is managed. Enucleation, local resection and radiotherapy are the most common therapeutic processes for the treatment of choroidal melanoma~\cite{p,p1,p2}.\\ Radiotherapy itself, depending on the source location, has two main classifications frequently regarded as internal and external source therapy. Brachytherapy, which is one of the most widely applied forms of treatment, belongs to the former treatment plan. From its initial appearance in the year 1930, brachytherapy, by the application of eye plaques, has played an indispensable role in treating patients suffering from various eye tumors in all around the world~\cite{d,e}. These plaques mainly function through conventional radionuclides such as $^{60}Co$, $^{106}Ru$, $^{125}I$, $^{103}Pd$, $^{90}Sr$ and $^{131}Cs$~\cite{f}. Admittedly, many state of the art plaques carry within themselves seeds of low-energy photon sources, while others merely operate as solid beta emitters~\cite{g}.\\
Considering the sensitive tissues which the eye is involved in, ocular tumors present a therapeutic challenge~\cite{h}. Although the radiation is effective therapeutically, it can be harmful
to healthy tissues. Plaque brachytherapy is the most widely used treatment which aims to transfer the maximum amount of dose to the tumor, while preventing dose absorption by normal tissue~\cite{i,i1}.
However, in the period of treatment, both the tumor and the proximal normal tissues obey a certain pattern in radiation absorption; hence, in this therapeutic procedure, minimizing the absorbed dose by the normal tissue is still one of the major concerns.\\ Applying the gold nanoparticles (GNPs) as radiation dose enhancer in combination with brachytherapy can be an effective method to reduce the radiation effects on healthy nearby tissues.\\ In the in-vivo study by Hainfeld et al., on the use of GNPs in mice which were irradiated by X-ray photons, the results show that the presence of GNPs in the tumor will cause more radiation dose in the cancerous cells than that of in the healthy tissues~\cite{j,j1}.\\ In the in-vitro study by T. Kong et al., on the application of GNPs to enhance radiation cytotoxicity, the results show that radiotherapy killed more cancerous cells in the presence of GNPs than that of in the absence of these nanoparticles~\cite{k}. Subsequently, GNPs radiosensitization has been observed in more controlled in-vitro irradiation of cells and plasmid DNA~\cite{l,l1,l2}.\\ The idea of using gold nanoparticles in cancer therapy as a radio-sensitizer is not a new one and several Monte Carlo, in-vitro and in-vivo studies on the application of nanotechnology-based cancer therapy have been performed using GNPs~\cite{x,x1,x2,x3}. However, only a handful of studies such as~\cite{m} have yet been conducted to investigate the effects of these nanoparticles on human eye tumors such as choroidal melanoma.\\ If the choroidal melanoma tumor could be loaded with these nanoparticles in proper concentrations and dimensions, this would lead to a higher absorbed dose by the tumor during a shorter time. Due to the fact that the eye is an extremely sensitive organ, the reduction of the period of treatment would decrease the absorbed dose by normal cells, resulting in a major cut down in the damage inflicted upon the mentioned organ.\\ Regarding the radio-sensitizing properties of gold nanoparticles, dose enhancement in the tumor is expected when the GNPs-loaded choroidal melanoma is locally irradiated with brachytherapy sources. However, considering the healthy tissues, and due to different seed models which are used in the eye plaque therapy, the variation of energy deposition in normal tissues, in comparison with the resultant dose increase in the tumor by diverse sources is of outmost importance in the investigation of GNPs effects on ophthalmic brachytherapy dosimetry.\\ Some studies have been carried out through making a dosimetry comparison between $^{103}Pd$ and $^{125}I$ ~\cite{n,g}. These studies reported higher absorbed dose by the tumor for $^{103}Pd$ versus $^{125}I$ for an equivalent radiation time. However, the focus of the present study is to know what changes will occur in absorbed dose by the healthy tissues compared to the dose increase in the tumor when the nanoparticles-induced tumor is irradiated with $^{125}I$ and $^{103}Pd$.\\ Here, in addition to providing a rigorous comparison between $^{103}Pd$ and $^{125}I$ low-energy photon sources within COMS eye plaques in the study of the effects of GNPs in ophthalmic brachytherapy, once again, we also studied the difference between water phantom and human eye globe in the presence and absence of these nanoparticles for both mentioned sources. The mean absorbed dose to the apex of the tumor as well as other critical points in both water and eye phantoms were calculated in order to evaluate the effects of both sources and identify the most efficient source with the least amount of induced cell damage to critical points and healthy tissue in the presence of GNPs.\\ To this end the eye globe was simulated following the way of previous work~\cite{m} precisely by considering different parts of the eye and their components. Here, both water and eye phantoms were simulated with MCNP5 code in which the water and eye phantoms were nearly identical, owing to the fact of water being taken as the composition of the simulated eye globe, albeit by considering accurate geometry of the noted organ. The geometric characteristic of this phantom is in accordance with the eye of an adult. A certain diameter of 50 nm was chosen for GNPs which has been shown to be the optimal size to achieve maximal uptake in mammalian cells~\cite{q}. Dosimetric characteristics of a single source for $^{125}I$ and $^{103}Pd$ were utilized to validate the accuracy of the Monte-Carlo simulation technique. Fully-loaded 16mm COMS standard eye-plaque with $^{125}I$ source model 6711 (GE Healthcare/Oncura) and also $^{103}Pd$ source model 200 (Theraseed) were the focus of this study.
\vspace*{2.2cm}
\section{Material and Methods}

The primary goal of this work is to examine and compare the efficiency of the use of ophthalmic brachytherapy in combination with GNPs in the treatment of ocular tumor between $^{125}I$ and $^{103}Pd$ sources.\\ The present Monte Carlo simulations are carried out using MCNP5 code~\cite{o6}. The mentioned code benefits from a three-dimensional heterogenous geometry for both photons and electrons situated within the energy range of 1 KeV to 1 GeV; moreover, the libraries incorporated within MCNP5 are based upon the 8th release of ENDF/B-VI~\cite{o7}. *F8 tally following $10^9$ histories and F6 tally following $10^7$ histories were used in this study to perform the phantom dosimetry and the air kerma simulations respectively, in order to achieve a relative statistical error of less than 1\% .\\ The current Monte-Carlo study was conducted with two phantom test cases (water phantom and complete simulated human eye) in which,
the ophthalmic brachytherapy dosimetry was evaluated in both cases, in the presence and the absence of gold nanoparticles using the full loaded eye plaque containing $^{125}I$ and $^{103}Pd$ sources. We simulated the human eye globe in a manner similar as indicated in our previous work~\cite{m}.\\ In order to define complete human eye globe, its components have been simulated using interface among different shapes with specific geometry and characteristics. The exact method of simulation of the eye globe and the tumor has been thoroughly explained in the previous report~\cite{m}; hence, all the details relating to the simulated phantoms are based upon the mentioned report; furthermore, in addition to the 16mm COMS eye plaque loaded with 13 $^{125}I$ seeds, a 16mm eye plaque containing 13 $^{103}Pd$ seeds with the exact same coordinates were defined. The center of the simulated eye globe coincided with the origin.\\ Rows of cubic voxels, with individual volumes of $0.05cm^3$ , were utilized to evaluate the depth dose of the $^{125}I$ and $^{103}Pd$ radionuclide sources within both water and eye phantoms. The mentioned voxels were assumed to be arranged on the central axis of the eye plaque, which has been described in previous study~\cite{m}.
 According to the work done in the previous study~\cite{m}, an equator temporal eye Melanoma with a height of 0.5cm is considered. The COMS-style eye plaque with the diameter of 16mm was modeled on the equator temporal periphery to the eyeball.\\
It is a scientifically proven fact that the optimum GNPs uptake within cancerous cells occurs when the noted substances are roughly 50nm in diameter~\cite{q}; hence, the study benefits from the deployment of 50nm GNPs within the tumor with varying concentrations of 7mg, 10mg, 18mg, 30mg. Given the noted facts, the tumor is latticed by identical cubes, each with a volume of $0.1cm^3$.\\ The dose enhancement factor(DEF) has been calculated to compare the effects of the presence of GNPs in brachytherapy for the noted configurations of sources and phantoms. DEF is the ratio of the absorbed dose by the tumor which is loaded with GNPs to the absorbed dose by the tumor with out these nanoparticles. Also, the dose distribution in different points of the eye has been determined for two mention sources in both eye and water phantoms. validity of dosimetry computations for both the simulated sources were investigated by parameterized calculations such as Air Kerma Strength, Dose Rate Constant, and Radial Dose Function(RDF), which were then compared with the reported results of Thomson et al and Rivard et al~\cite{o,o1}. For RDF calculations, in accordance with previous studies~\cite{m}, each of the brachytherapy sources were simulated within a water phantom, $30cm^3$ in volume; furthermore, for the calculations related to Dose Fall-Off, tori cells with radial distances of 0.5cm to 10cm were simulated.
\vspace*{3cm}
\renewcommand{\thesubsection}{\Alph{subsection}}
\section{Results}
\vspace*{.2cm}
\subsection{Calculations of TG-43 parameter}
\begin{figure}[h]
      \centerline{\includegraphics[width=10cm]{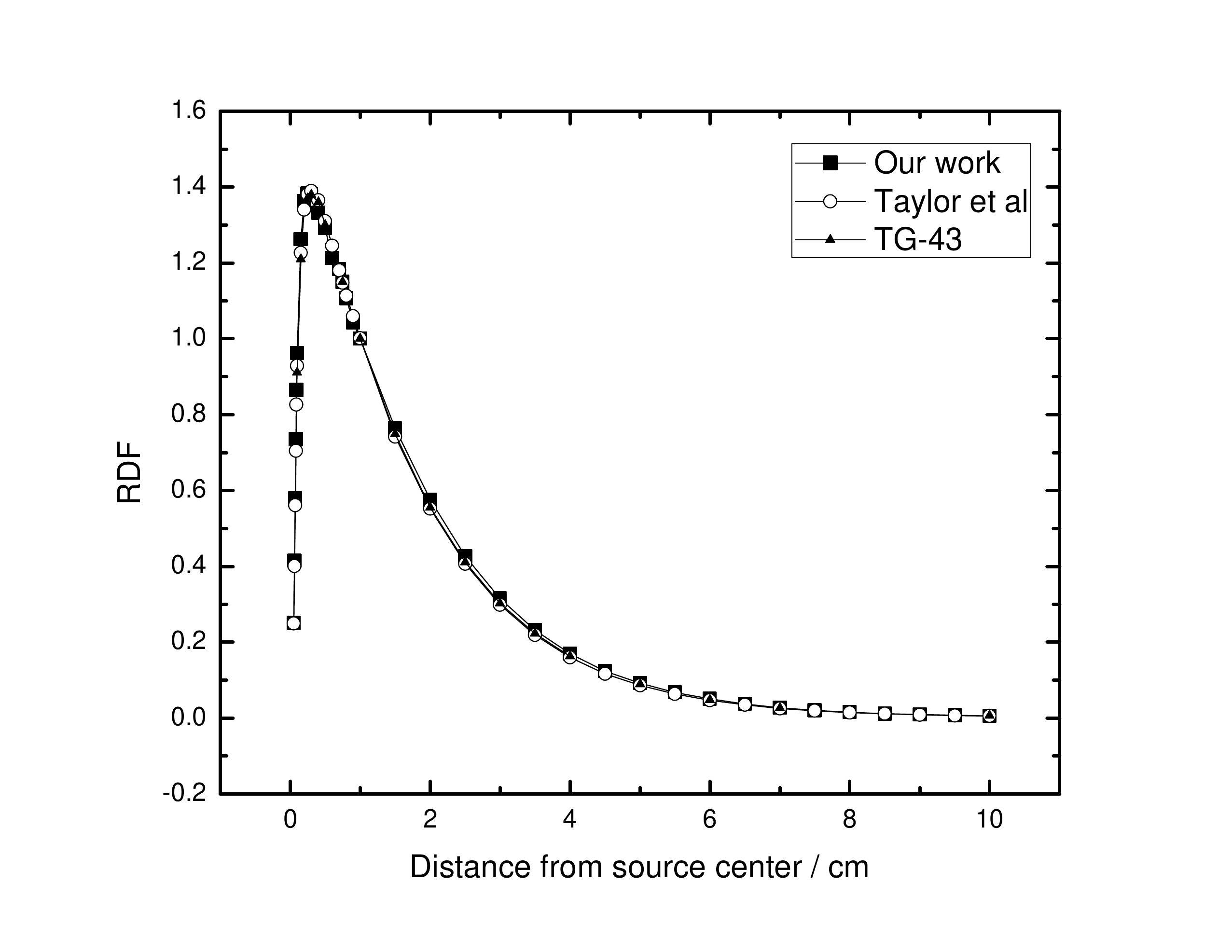}}
      \vspace*{-.7cm}
  \caption{The calculated Radial dose function for $^{103}Pd$ source. The relative statistical uncertainties is less than 1\%. The size of voxeles which the dose was scored in are the same as the previous work~\cite{m}.}\label{f-1}
\end{figure}

The TG-43 dosimetry parameter (Air Kerma Strength, Dose rate constant and RDF) have been calculated for $^{103}Pd$ following the approach of the previous work~\cite{m}. The calculated RDF in the present work, those reported by Thomson et al~\cite{o} and TG-43~\cite{o1} have been shown in Table\ref{Tab-1} and Fig.\ref{f-1} . Also, the calculated dose rate constant in this work, those reported by Thomson et al and TG-43 have been shown in Table\ref{Tab-2}. The results show excellent agreement between them. It is also noteworthy to mention that all the referred TG-43 parameter dosimetry were calculated for $^{125}I$ in the previous work~\cite{m}.
\vspace*{-.5cm}
\vspace*{1.7cm}
\subsection{Dosimetry Calculations}
\begin{figure}
  \centerline{\includegraphics[width=10cm]{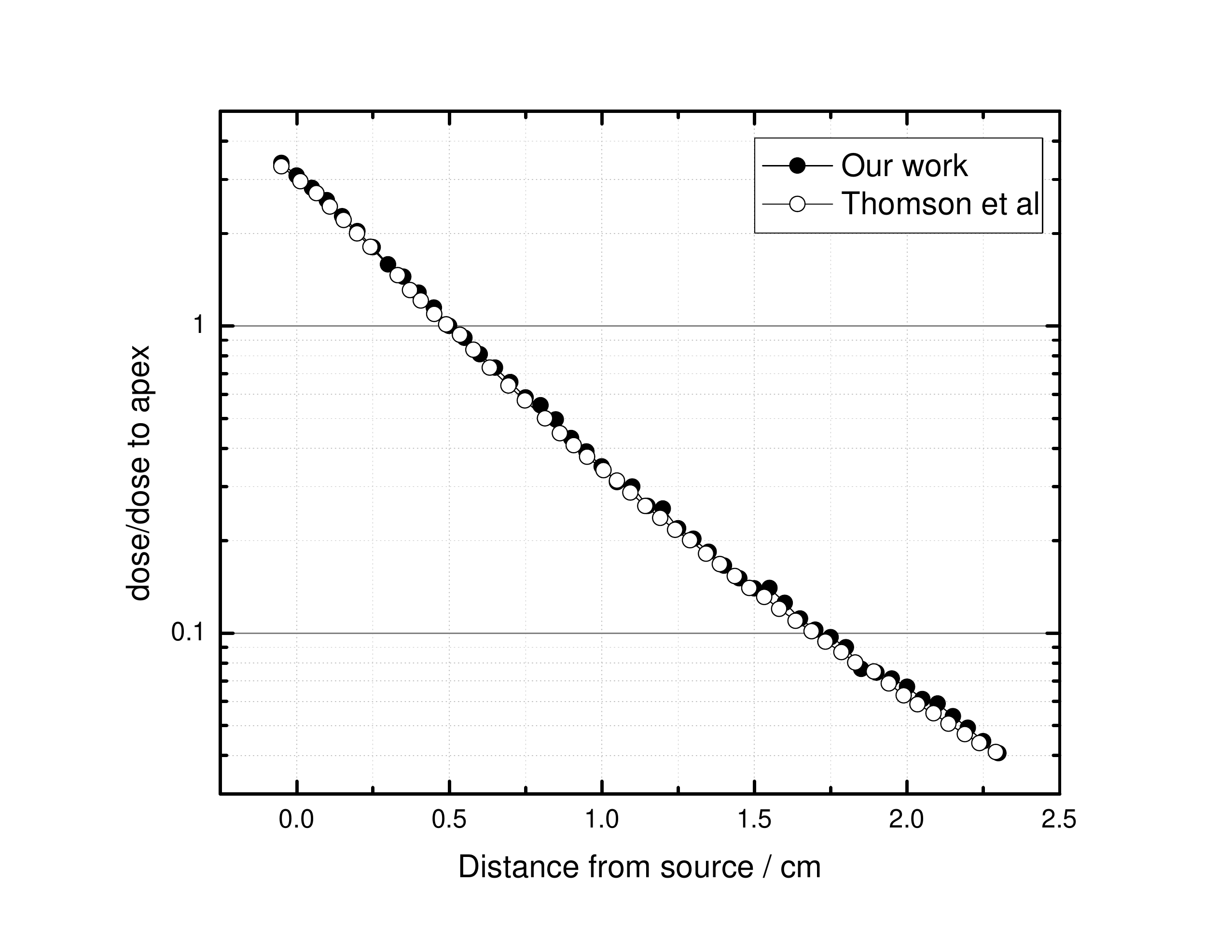}}
  \vspace*{-.7cm}
  \caption{The ratio of plaque central axes to the dose at the tumor apex for $^{103}Pd$ in the water phantom. The results have been compared with Thomson et al~\cite{o}.}\label{f-5}
\end{figure}

\begin{figure}[h!]
  \centerline{\includegraphics[width=10cm]{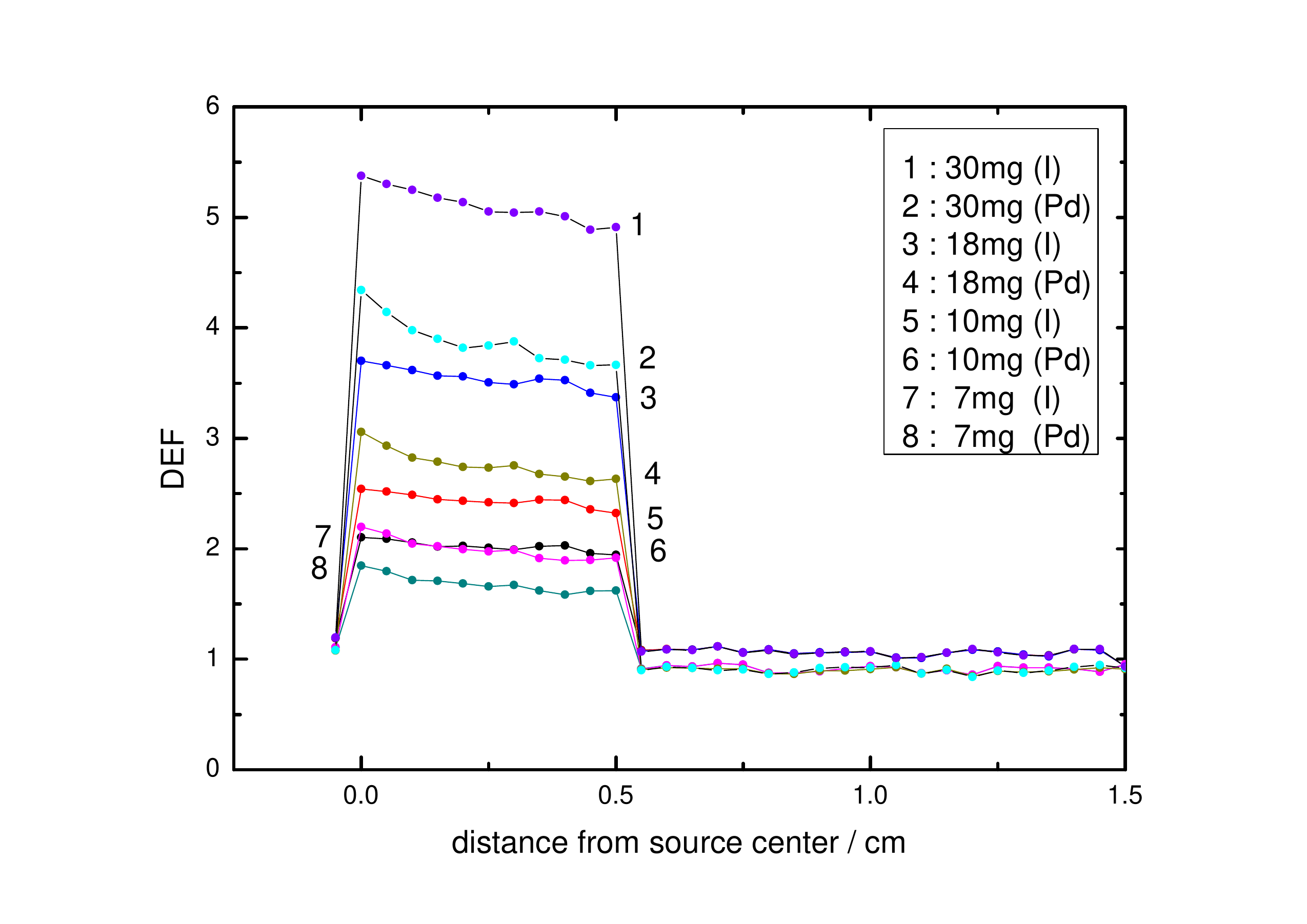}}
  % Requires \usepackage{graphicx
  \vspace*{-.7cm}
  \caption{The calculated (DEF) for 50nm GNPs within the tumor with concentrations of(7, 10, 18 and 30mg)/(g of tumor). The calculation have been done in the water phantom with fully-loaded 16mm COMS eye plaque of $^{125}I$ and $^{103}Pd$ sources. }\label{f-3}
\end{figure}

\begin{figure}[h!]
  \centerline{\includegraphics[width=10cm]{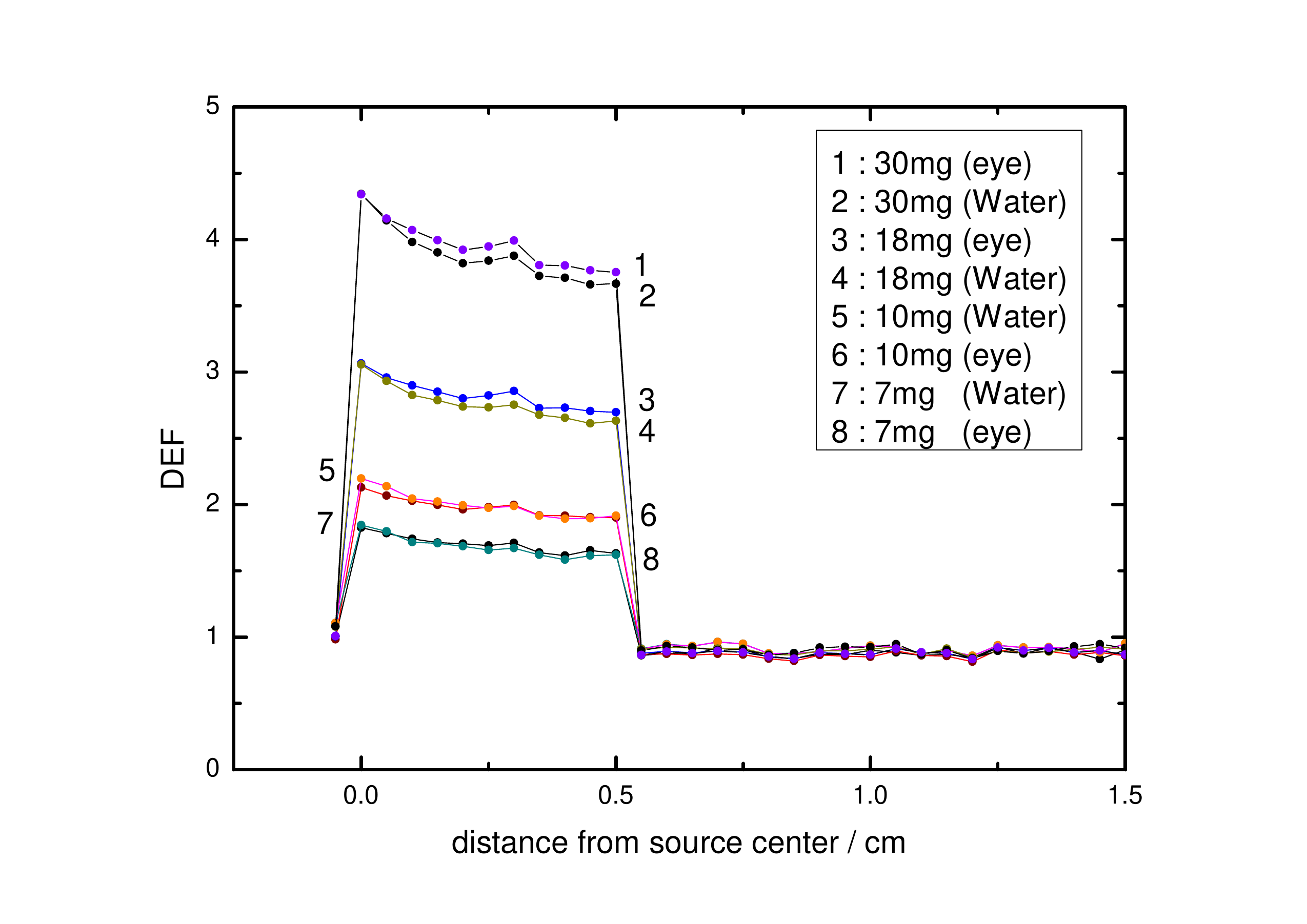}}
  \vspace*{-.7cm}
  \caption{The calculated (DEF) for 50nm GNPs within the tumor with concentrations of(7, 10, 18 and 30mg)/(g of tumor) in the eye and water phantoms for $^{103}Pd$ source.}\label{f-4}
\end{figure}

Fig.\ref{f-5} shows the depth dose curve in the plaque central axis direction for $^{103}Pd$ source in the fully-loaded 16mm COMS eye plaque in water phantom where the dose is presented relative to the dose at the tumor apex. The calculations have been done by *F8 tally. The results have been compared with those reported by Thomson et al~\cite{o} with an excellent agrement between them.
\vspace*{-.11cm}
\\
\\
\\
\\
\\
\\
\\
\\
\\
\\
The dose to points of interest has been calculated in the water phantom and eye globe for both $^{125}I$ and $^{103}Pd$ sources and the results have been reported in Table\ref{Tab-3}. Having yielded a relative statistical uncertainty is less than 1\%, with the maximum percentage being apparent in the opposite side of the eye, and the minimal amount of the noted factor observable in the sclera, the study has tabulated the full body of prescription points within the aforementioned Table\ref{Tab-3}.\\The dose enhancement factor (DEF) has been calculated for both $^{125}I$ and $^{103}Pd$ sources. This calculation has been done for different concentration of GNPs within the water phantom and compared with the calculated DEF for $^{125}I$ in Fig.\ref{f-3}.
\\A comparison between water phantom and eye globe in calculation of dose enhancement factor for $^{103}Pd$ is shown in Fig.\ref{f-4}. The tally cells have been placed along the central axis of the plaque, starting from the sclera near the plaque to the sclera opposite side of the plaque.

\vspace*{1cm}
\section{Discussion}
Regarding the radiation enhancement properties of GNPs, studies have indicated that in radiotherapy of cancerous cells, the presence of these nanoparticles in the tumor site will locally increase the absorbed dose. Delivering the required dose in a shorter period of time will ultimately result in reduced time of radiation exposure, which in turn would diminish the delivered dose to the normal tissues.\\
In the previous study~\cite{m}, the dose enhancement properties of GNPs on Choroidal Melanoma for $^{125}I$ source were investigated.
Given the differences in the properties of ophthalmic brachytherapy sources, the effects of the presence of GNPs in the tumor site in minimizing the damage done to normal tissue for each of these sources can be further investigated and compared.\\
The results of this study have been quick to point out that, in brachytherapy of Choroidal Melanoma, the $^{125}I$ source has granted us a greater DEF in the presence of GNPs when compared with the $^{103}Pd$ source.
As it is pointed out in Fig.\ref{f-3}, the resultant DEF in the apex of the tumor for GNPs concentrations of 30, 18, 10 and 7 mg/g are equal to 4.91, 3.37, 2.32, 1.94 respectively for $^{125}I$ source , while are 3.66, 2.63, 1.91 and 1.62 respectively for the $^{103}Pd$ source. The apparent deviances for the mentioned sources are greater for higher concentration.\\
It is also noteworthy to mention that, the dosimetry calculations for the eye in the presence of GNPs report no eye-catching differences for the absorbed dose in the healthy tissue in comparison with the case where GNPs were absent for each of the noted sources.
In reference to the noted evidence, in brachytherapy of eye Melanoma in presence of GNPs, it is more efficient to utilize the $^{125}I$ source, for it grants a higher DEF when compared with the $^{103}Pd$ source.\\
The importance of defining an actual eye phantom instead of the water phantom in Monte Carlo studies has been another factor gaining much attention in the recent years.
The results of Fig.\ref{f-4} signify that calculations of the resultant DEF in the tumor for the brachytherapy of eye Melanoma utilizing a $^{103}Pd$ source show no striking difference when the actual eye model and the water phantom are compared. This is while according to the previous study~\cite{m}, the results pointed out that the use of the actual eye phantom in the Monte-Carlo study of the dosimetry in eye Melanoma, with $^{125}I$ source, whilst GNPs were present, was of utmost importance.\\
Furthermore, the results obtained from Table\ref{Tab-3} indicate that in a healthy tissue such as the lens, the the difference in the calculated absorbed dose, between water and eye phantoms, in varying concentrations of GNPs, is around 19\% for the $^{125}I$ source, and 14\% for the $^{103}Pd$ source. As stated previously, the deviances for both phantoms, in the presence of GNPs are greater when compared to those that were in the absence of the mentioned substance.\\
As it is apparent from Table\ref{Tab-3}, the calculated difference in the DEF at the apex of the tumor, between water and eye phantoms for the $^{125}I$ source is roughly 4\%, while the noted deviance is about 2\% for the $^{103}Pd$ source.\\
All in all, regarding the numerous sources that are employed in the radiotherapy of the eye Melanoma, the possible effects of these sources, whilst accompanied by GNPs, in altering the period of treatment, is a matter of heated debate, and requires more comprehensive investigations.

\vspace*{0.1cm}
\section{Conclusion}
The present study indicates that the presence of GNPs as a method of dose enhancement in treatment of eye tumors grants a higher DEF for the $^{125}I$ brachytherapy source as opposed to that of the $^{103}Pd$ source.\\
 Previous studies have been quick to point out that, in the period of brachytherapy treatment of the eye(in the absence of GNPs) the $^{103}Pd$ source generates a greater dose in the tumor site. Since the results designate that the difference between the absorbed dose in normal tissue, in the presence of GNPs, is negligible when the two sources are compared, it is safe to say that the use of the $^{125}I$ source along with GNPs would yield a higher DEF in the tumor site of the eye Melanoma.\\
  The previous study has also specified that ever-more accurate definition of the actual eye phantom, instead of the water phantom, is an absolute necessity for precise dosimetry of the eye Melanoma when examination utilizing the $^{125}I$ source is intended. However, even with the aforementioned facts, the dosimetry results obtained for the $^{103}Pd$ source in both the water and eye phantoms are roughly similar; thus, in brachytherapy studies of the eye involving Monte-Carlo methods, designation of a water phantom as an alternative for the eye phantom breeds no observable difference in the results, and efficiently replicates those of the eye phantom.
\vspace*{1.4cm}
\section{Tables}
\begin{table}[h!]
\centering
\vspace*{-.1cm}
\caption{A comparison between the simulated (RDF) of $^{103}Pd$ source in this work with the published data by other investigators.}
\label{t-2}
\vspace*{.1cm}
\begin{tabular}{|c|c|c|c|}
  \hline
  \multicolumn{4}{|c|}{Radial Dose Function, g(r )} \\
  \hline
  Distance from source(cm) & This work & Thomson et al~\cite{o} & TG-43~\cite{o1} \\
  \hline
  0.05 & 0.25067 & 0.249 & --- \\
  \hline
  0.06 & 0.41397 & 0.401 & --- \\
  \hline
  0.07 & 0.57996 & 0.561 & --- \\
  \hline
  0.08 & 0.73469 & 0.704 & --- \\
  \hline
  0.09 & 0.86419 & 0.826 & --- \\
  \hline
  0.1 & 0.96223 & 0.929 & 0.911 \\
  \hline
  0.15 & 1.26322 & 1.226 & 1.21 \\
  \hline
  0.2 & 1.36332 & 1.34 & --- \\
  \hline
  0.25 & 1.38322 & 1.381 & 1.37 \\
  \hline
  0.3 & 1.38156 & 1.39 & 1.38 \\
  \hline
  0.4 & 1.33207 & 1.365 & 1.36 \\
  \hline
  0.5 & 1.29172 & 1.31 & 1.3 \\
  \hline
  0.6 & 1.21264 & 1.245 & --- \\
  \hline
  0.7 & 1.18272 & 1.18 & --- \\
  \hline
  0.75 & 1.14951 & 1.148 & 1.15 \\
  \hline
  0.8 & 1.10612 & 1.113 & --- \\
  \hline
  0.9 & 1.04254 & 1.059 & --- \\
  \hline
  1 & 1.00059 & 1.001 & 1 \\
  \hline
  1.5 & 0.76306 & 0.742 & 0.749 \\
  \hline
  2 & 0.57565 & 0.552 & 0.555 \\
  \hline
  2.5 & 0.42646 & 0.407 & 0.41 \\
  \hline
  3 & 0.31547 & 0.298 & 0.302 \\
  \hline
  3.5 & 0.23145 & 0.219 & 0.223 \\
  \hline
  4 & 0.16997 & 0.16 & 0.163 \\
  \hline
  4.5 & 0.12385 & 0.117 & --- \\
  \hline
  5 & 0.09241 & 0.0865 & 0.0887 \\
  \hline
  5.5 & 0.06699 & 0.0635 & --- \\
  \hline
  6 & 0.05031 & 0.0469 & 0.0482 \\
  \hline
  6.5 & 0.03689 & 0.0346 & --- \\
  \hline
  7 & 0.02696 & 0.0256 & 0.0262 \\
  \hline
  7.5 & 0.02069 & 0.0193 & --- \\
  \hline
  8 & 0.01552 & 0.0147 & --- \\
  \hline
  8.5 & 0.01175 & 0.0112 & --- \\
  \hline
  9 & 0.00893 & 0.0084 & --- \\
  \hline
  9.5 & 0.00703 & 0.0064 & --- \\
  \hline
  10 & 0.00523 & 0.0051 & 0.00615 \\
  \hline
\end{tabular}
\label{Tab-1}
\end{table}

\vspace*{0.5cm}

\begin{table}[h!]
\centering
\caption{ A comparison of the dose rate constant of $^{103}Pd$ brachytherapy source in water, simulated in this project with the published data.}
\label{t-1}

\begin{tabular}{|c|c|c|c|}
  \hline
   & Thomson et al~\cite{o} & M.J. Rivard et al~\cite{o1} & This work \\
  \hline
  Dose rate constant & 0.772 & 0.686 & $ 0.69 \pm 0.01 $ \\
  \hline
\end{tabular}
\label{Tab-2}
\end{table}

\begin{table}[!]
\centering
\caption{ A comparison of the dose (unit of (Gy/per particle) $\times10^{-15}$) to the critical points of the eye in the water and eye phantoms for both fully-loaded 16mm (13-seeds) COMS standard $^{125}I$ and $^{103}Pd$ eye plaques. Eye refers to the eye phantom and water refers to the water phantom in which the eye phantom was filled of water. `` 7mg/g, 10mg/g, 18mg/g and 30mg/g" refer to the concentration of GNPs inside the tumor.}
\label{t-3}
\vspace*{.1cm}
\scalebox{0.75}{

\begin{tabular}{|c|c|c|c|c|c|c|c|c|c|c|}
  \hline
  \multicolumn{11}{|c|}{Water} \\
  \hline
   & $^{125}I$  & $^{103}Pd$  & 7mg/g  & 7mg/g  & 10mg/g  & 10mg/g  & 18mg/g  & 18mg/g & 30mg/g & 30mg/g  \\
  Location &  &  & ($^{125}I$) & ($^{103}Pd$) & ($^{125}I$) & ($^{103}Pd$) & ($^{125}I$) & ($^{103}Pd$) & ($^{125}I$) & ($^{103}Pd$) \\
  \hline
  Sclera & 138.97 & 217.97 & 165.45 & 241.12 & 165.56 & 241.00 & 165.86 & 235.10 & 166.16 & 235.86 \\
  \hline
  apex & 41.03 & 64.31 & 79.78 & 104.18 & 95.22 & 123.00 & 138.29 & 169.24 & 201.48 & 235.67 \\
  \hline
  Center of eye & 13.85 & 19.28 & 14.09 & 16.86 & 14.01 & 16.80 & 14.01 & 16.78 & 14.10 & 16.76 \\
  \hline
  Opposite side & 2.91 & 2.62 & 3.22 & 2.12 & 3.20 & 2.13 & 3.20 & 2.37 & 3.24 & 2.48 \\
  \hline
  Optic nerve & 4.99 & 5.05 & 5.07 & 4.59 & 5.07 & 4.59 & 5.07 & 4.74 & 5.08 & 4.93 \\
  \hline
  Lens & 9.54 & 11.92 & 10.72 & 11.41 & 10.77 & 11.40 & 10.77 & 11.34 & 10.73 & 11.39 \\
  \hline
  Macula & 6.97 & 7.68 & 7.72 & 7.48 & 7.71 & 7.48 & 7.71 & 7.52 & 7.77 & 7.47 \\
  \hline
\end{tabular}
\label{Tab-3}
}
\end{table}

\scalebox{0.75}{

\begin{tabular}[h!]{|c|c|c|c|c|c|c|c|c|c|c|}
  \hline
  \multicolumn{11}{|c|}{eye} \\
  \hline
   & $^{125}I$  & $^{103}Pd$  & 7mg/g  & 7mg/g  & 10mg/g  & 10mg/g  & 18mg/g  & 18mg/g & 30mg/g & 30mg/g  \\
  Location &  &  & ($^{125}I$) & ($^{103}Pd$) & ($^{125}I$) & ($^{103}Pd$) & ($^{125}I$) & ($^{103}Pd$) & ($^{125}I$) & ($^{103}Pd$) \\
  \hline
  Sclera & 174.89 & 263.62 & 184.90 & 265.00 & 185.18 & 259.47 & 185.53 & 265.26 & 185.81 & 265.95 \\
  \hline
  apex & 43.12 & 64.37 & 80.50 & 105.00 & 96.21 & 122.39 & 138.46 & 173.59 & 200.69 & 241.61 \\
  \hline
  Center of eye & 14.76 & 18.82 & 13.46 & 16.20 & 13.48 & 16.25 & 13.47 & 16.67 & 13.53 & 16.66 \\
  \hline
  Opposite side & 3.75 & 2.89 & 3.41 & 2.41 & 3.39 & 2.95 & 3.39 & 2.69 & 3.38 & 2.69 \\
  \hline
  Optic nerve & 4.81 & 4.95 & 4.96 & 4.42 & 4.97 & 4.48 & 4.97 & 4.60 & 4.96 & 4.59 \\
  \hline
  Lens & 8.55 & 10.23 & 8.74 & 9.71 & 8.65 & 9.71 & 8.65 & 9.73 & 8.62 & 9.73 \\
  \hline
  Macula & 7.14 & 7.24 & 7.61 & 7.11 & 7.64 & 7.39 & 7.64 & 7.41 & 7.65 & 7.40 \\
  \hline
\end{tabular}
}

\vspace*{9cm}
‎

\end{document}